\documentclass[conference]{IEEEtran}
\IEEEoverridecommandlockouts
\usepackage{cite}
\usepackage{amsmath,amssymb,amsfonts}
\usepackage{algorithmic}
\usepackage{graphicx}
\usepackage{textcomp}
\usepackage{xcolor}
\usepackage{booktabs}
\usepackage{pifont}
\usepackage{multirow}
\usepackage{adjustbox}
\usepackage{hyperref}
\def\BibTeX{{\rm B\kern-.05em{\sc i\kern-.025em b}\kern-.08em
    T\kern-.1667em\lower.7ex\hbox{E}\kern-.125emX}}
\begin{document}

\title{FoleyGRAM: Video-to-Audio Generation with GRAM-Aligned Multimodal Encoders\\
\thanks{* Equal contribution.\\Corresponding author's email: \text{riccardofosco.gramaccioni@uniroma1.it}}
}

%\author{Riccardo~F.~Gramaccioni$^{\flat}$, Christian~Marinoni$^{\flat}$, Kazuki~Shimada$^{\sharp}$,\\Takashi~Tak~Shibuya$^{\sharp}$, Yuki~Mitsufuji$^{\sharp \natural}$, and~Danilo~Comminiello$^{\flat}$ \\\\\textit{$^{\flat}$Sapienza University of Rome, Italy}\\\textit{$^{\sharp}$Sony AI, Tokyo, Japan}\\\textit{$^{\natural}$Sony Group Corporation, Tokyo, Japan}
%}

% \author{\IEEEauthorblockN{Anonymous Authors}}
%  QUESTO é giusto
\author{\IEEEauthorblockN{Riccardo~F.~Gramaccioni$^*$, Christian~Marinoni$^*$, Eleonora~Grassucci, \\Giordano~Cicchetti, Aurelio~Uncini, and Danilo~Comminiello}\\
        % <-this % stops a space
        \IEEEauthorblockN{\textit{Dept. Information Engineering, Electronics and Telecommunications (DIET), Sapienza University of Rome, Italy}}
}

\maketitle

\begin{abstract}
In this work, we present FoleyGRAM, a novel approach to video-to-audio generation that emphasizes semantic conditioning through the use of aligned multimodal encoders. Building on prior advancements in video-to-audio generation, FoleyGRAM leverages the Gramian Representation Alignment Measure (GRAM) to align embeddings across video, text, and audio modalities, enabling precise semantic control over the audio generation process. The core of FoleyGRAM is a diffusion-based audio synthesis model conditioned on GRAM-aligned embeddings and waveform envelopes, ensuring both semantic richness and temporal alignment with the corresponding input video. We evaluate FoleyGRAM on the Greatest Hits dataset, a standard benchmark for video-to-audio models. Our experiments demonstrate that aligning multimodal encoders using GRAM enhances the system's ability to semantically align generated audio with video content, advancing the state of the art in video-to-audio synthesis. 
%Code and samples are available at our demo page.
\end{abstract}


\section{Introduction}
In recent years, transforming visual information into audio representations, known as video-to-audio (V2A) generation task, has gained increasing attention. V2A task is discovering extremely attractive applications in fields concerning sound design in cinema and video games, enhancing accessibility tools, and creating immersive multimedia experiences. Central to this challenge is the ability to generate audio that not only matches the temporal and structural properties of the visual input but also captures its semantics.

Usually, multiple semantic inputs can be used in the process of generating audio, as different semantic conditioning may allow the control of diverse aspects of the generated waveform \cite{Wu2023MusicCM}. Typically, for V2A task, semantics is controlled through video, audio, or text conditioning, and existing methods rely on encoder architectures to condition the audio generation process on such relevant visual and semantic cues \cite{Comunit2023SyncfusionMO, Zhou2017VisualTS, Chen2024SemanticallyCV, Cui2022VarietySoundTV, 9782577}. While effective in some cases, this approach has severe limitations that undermine the effective control of semantics in generated audio. A significant limitation of these approaches lies in the lack of joint training for the encoders used across different modalities. This disjoint training paradigm often results in the creation of separate latent spaces for each modality, leading to misaligned embeddings that the generative model may semantically badly interpret \cite{moschella2023relative}. Additionally, even in the case of jointly-trained encoders, misalignment in the latent space may occur, as all previous methods solely rely on cosine similarities that can only be computed between pairs of modalities \cite{saporta2024contrasting}. More specifically, state-of-the-art models select an anchor modality and align all other modalities to the anchor. Examples are ImageBind \cite{Girdhar2023ImageBindOE} that selects the image modality as anchor, or LanguageBind \cite{Zhu2023LanguageBindEV}, selecting the text modality instead. Although promising, this approach does not provide any geometrical guarantees that the other modalities are aligned with each other and, in practice, they are not \cite{saporta2024contrasting, cicchetti2024GRAM}. Therefore, during training, such encoders may end up in a local minimum or may not guarantee all the modalities' true geometric alignment together. Such misalignment can compromise the semantic coherence of the generated audio, reducing the model's ability to faithfully represent the desired audiovisual relationship.

\begin{figure}[t!]
    \centering
    \includegraphics[width=\linewidth]{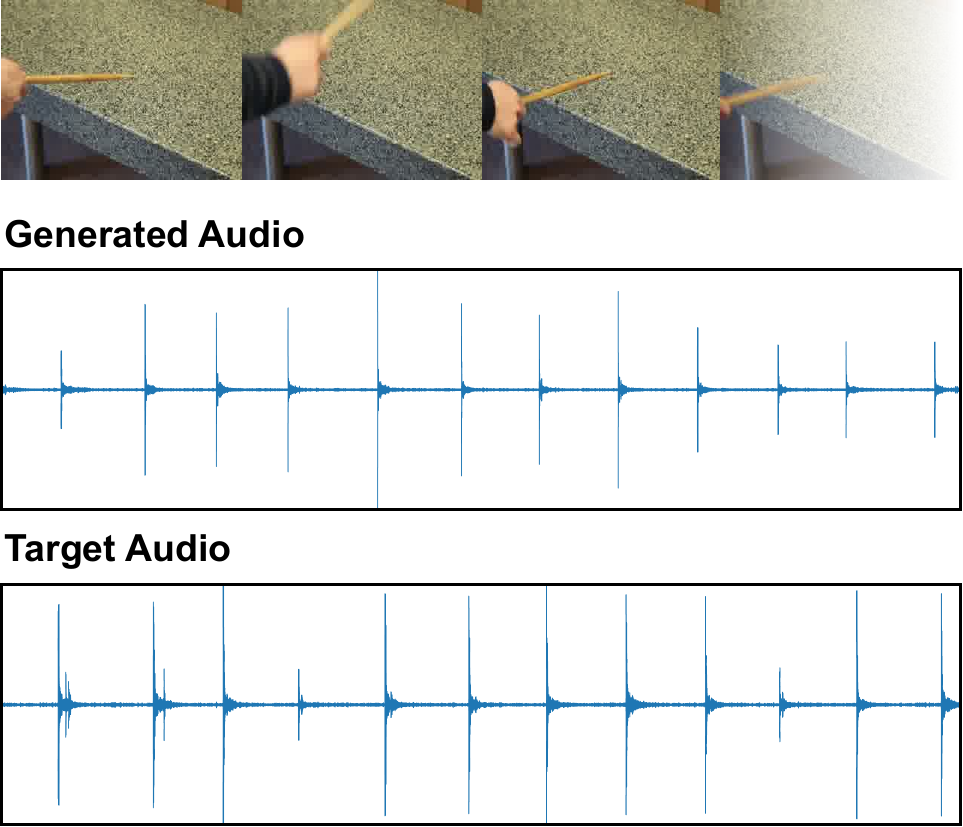}
    \caption{Example showing ground truth audio and video and relative waveform generated by the proposed method.}
    \label{fig:risultati-spettrogrammi}
\end{figure}

To address this limitation, we propose FoleyGRAM, a novel approach that leverages the Gramian Representation Alignment Measure (GRAM) \cite{cicchetti2024GRAM} to ensure aligned latent representations across multiple modalities. GRAM enables the construction of a shared latent space that is jointly trained and optimized, providing a robust framework for embedding alignment. Indeed, GRAM relies on the computation of the volume of the high-dimensional parallelotope defined by the modalities embeddings, which provides direct insights into the joint alignment of all the modalities at once, avoiding pairwise computations. By aligning the latent spaces of video, text, and audio modalities, FoleyGRAM facilitates precise semantic conditioning, enhancing the quality and relevance of the generated audio.
At the generative core of FoleyGRAM is a diffusion-based audio synthesis model, conditioned on GRAM-aligned embeddings and additional waveform envelope information. This dual conditioning mechanism ensures semantic fidelity through GRAM and also temporal synchronization between the input video and the generated audio by means of the envelope. The effectiveness of our approach is demonstrated on the Greatest Hits dataset, a benchmark for video-to-audio generation. Experimental results show that FoleyGRAM achieves superior results compared with common baseline methods for V2A tasks, with better semantic alignment and audio quality. An example of a result is shown in Fig.~\ref{fig:risultati-spettrogrammi}.

Our main contributions can be summarized as follows:

\begin{itemize}

\item We propose FoleyGRAM, a novel V2A model able to generate semantically meaningful and temporally aligned audio from video.

\item We use GRAM for producing highly semantically aligned embeddings for the generative model conditioning, resulting in unified semantic controls.

\item FoleyGRAM achieves enhanced semantic fidelity through the use of such unified, jointly trained and optimized latent space. Comprehensive evaluations validate the effectiveness of our approach, demonstrating advancements in semantic alignment and generative quality.

\end{itemize}

Through these contributions, FoleyGRAM represents a significant step forward in video-to-audio generation, offering new solutions for multimodal semantic conditioning in generative models.

The rest of the paper is organized as follows. Section~\ref{sec:works} presents the related works, Section~\ref{sec:method} the proposed method, while in Section~\ref{sec:exp} we discuss the experimental results and in Section~\ref{sec:res} we validate the obtained results. Finally, conclusions are drawn in Section~\ref{sec:con}.

\section{Related Works}
\label{sec:works}

\textbf{Video-to-Audio Generation.} The task of generating audio aligned with video has gained increasing attraction in multimedia post-production, driven by recent advancements in deep learning. Several state-of-the-art models have been proposed, aiming to achieve both semantic coherence and temporal alignment between the visual input and the generated audio. Early approaches, such as Im2Wav \cite{Sheffer2022IHY}, utilized transformer-based architectures conditioned on visual features extracted using CLIP \cite{Radford2021LearningTV}, while models like RegNet \cite{Chen2020GeneratingVA} employed GANs with video encoders to synthesize temporally aligned audio from video inputs. These efforts demonstrated the potential of multimodal learning but often suffered from limitations in alignment precision and semantic control.

Recent innovations, including SpecVQGAN \cite{SpecVQGAN_Iashin_2021} and Diff-Foley \cite{NEURIPS2023_98c50f47}, have further improved temporal and semantic alignment by leveraging optical flow features and contrastive learning strategies. For instance, Diff-Foley employs Contrastive Audio-Visual Pretraining (CAVP) to align video and audio embeddings before conditioning a latent diffusion model. Similarly, CondFoleyGen \cite{du2023conditional} demonstrates the utility of training directly on benchmark datasets, achieving improved alignment through Transformer-based architectures. However, these methods lack human-intelligible controls, limiting their utility in practical sound design applications \cite{gramaccioni2024folai}.
SyncFusion \cite{Comunit2023SyncfusionMO} addresses some of these challenges by introducing a human-readable control mechanism based on onset tracks. While this approach provides temporal guidance for audio generation, it requires manually annotated datasets and may not capture finer semantic details, such as sound intensity or duration. Finally, models like T-Foley \cite{Chung2024TFoleyAC} demonstrated the effectiveness of envelope-based conditioning for precise temporal alignment but lacked the flexibility to integrate semantic controls across multiple modalities.

\textbf{Multimodal Alignment.} The alignment of multiple modalities is a crucial and challenging task for enabling deep learning models to understand surrounding reality and generate content accordingly. The introduction of foundational models for two modalities like CLIP \cite{Radford2021LearningTV} for text and images has significantly influenced cross-modal alignment, inspiring subsequent works such as CLAP \cite{CLAP2022} for audio-text alignment. Such works rely on the cosine similarity between the two modalities and establish the conventional receipt for multimodal alignment. Indeed, the pairwise cosine similarity has been leveraged in following works like CLIP4VLA \cite{Ruan2023AccommodatingAM} integrating text, images, and audio samples, ImageBind \cite{Girdhar2023ImageBindOE}, LanguageBind \cite{Zhu2023LanguageBindEV}, and VAST \cite{Chen2023VASTAV} scaling up to 5 modalities. Despite the improved performance, these methods rely on the same cosine similarity loss function and align all the modalities to a select anchor one, providing no guarantees that all other modalities are aligned with each other, thus limiting the expressiveness of the latent space and resulting in modalities that may not be aligned in practice.

% The Gram matrix has emerged as a powerful tool for embedding alignment across diverse applications, including neural style transfer, sound event detection, and domain adaptation. Building on this foundation, multimodal models like CLIP4VLA and ImageBind have incorporated Gram-based measures to align embeddings across multiple modalities. Despite these advancements, existing models often rely on cosine similarity or simple fusion strategies, which fail to leverage the high-dimensional information required for complex multimodal tasks.

% SynGRAM, leveraging a state-of-the-art audio synthesis model introduces in Stable-V2A and based on Stable Audio, builds on these foundational works by introducing a unified, high-dimensional latent space for multimodal embedding alignment. Leveraging GRAM, SynGRAM ensures that video, audio, and text embeddings are semantically aligned, providing both precise control and enhanced generative quality. This approach addresses the limitations of prior models, offering new solutions for video-to-audio semantic synthesis.

\section{FoleyGRAM}
\label{sec:method}

\subsection{Gramian Representation Learning}

\begin{figure}
    \centering
    \includegraphics[width=\linewidth]{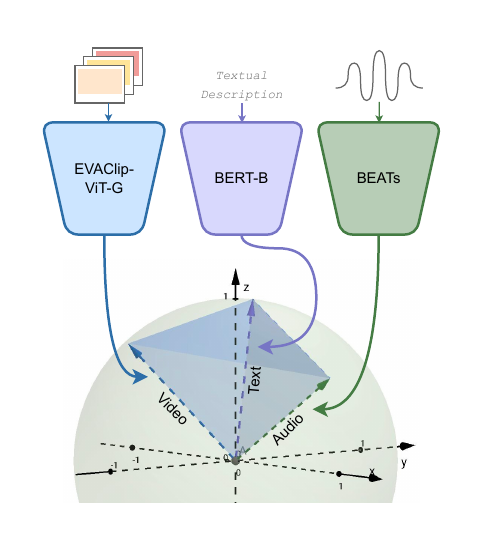}
    \caption{GRAM framework, in which the representation learned from the three encoders (EVAClip-ViT-G for video, BERT-B for text, and BEATs for audio) shape the edges of the high-dimensional parallelotope, whose volume provides insights on the alignment of the data.}
    \label{fig:gram}
\end{figure}

\begin{figure*}
    \centering
    \includegraphics[width=\linewidth]{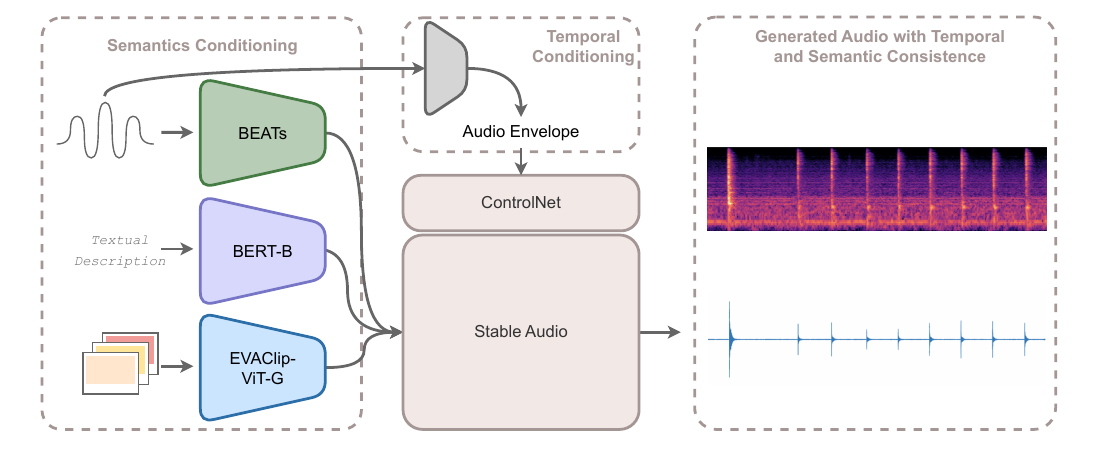}
    \caption{\textbf{FoleyGRAM architecture}: relevant semantic features are extracted from reference video, audio, and text through GRAM-aligned multimodal encoders. These features are used to condition an audio synthesis model that, together with the temporal information provided as an envelope signal used as input to a ControlNet, generates an audio that is temporally and semantically aligned with the reference video. At inference time, the three modalities can be used jointly or separately to generate the desired output. The samples used to condition the generation process can also be completely different from the semantic characteristics related to the video to be sonorized, allowing the sound designers to choose as they like the samples with which they can define the semantics for the audio to be generated.}
    \label{fig:architecture}
\end{figure*}

Conventionally, multimodal models align their representations according to the cosine similarity score between pairs of modalities. The cosine similarity is incorporated into the InfoNCE loss \cite{Oord2018RepresentationLW} as done for two modalities by CLIP \cite{Radford2021LearningTV}. However, when scaling to more than two modalities like in the video-audio-text case, the cosine similarity-based loss has severe limitations and fails to learn a unified latent space, obtaining suboptimal performance in downstream tasks \cite{cicchetti2024GRAM, saporta2024contrasting}.
To avoid such limitations, we involve GRAM \cite{cicchetti2024GRAM}, a recent multimodal model able to learn a unified latent space by means of a brand-new loss function. The GRAM loss function is based on the intuition that modalities embedding vectors lie in a hypersphere with unitary norm and that those vectors act as the edges of a high-dimensional parallelotope. Then, the volume of such parallelotope provides direct information about the alignment of the vectors, being small in the case of aligned data and large in the case of vectors representing different semantic concepts, as shown in Fig.~\ref{fig:gram}.
More formally, consider the three latent representations of audio $\mathbf{a}$, video $\mathbf{v}$, and text $\mathbf{t}$ be vectors in $\mathbb{R}^n$ arranged in a matrix $\mathbf{A}$ containing its dot products. From $\mathbf{A}$ we can easily compute the Gram matrix as $\textbf{G}(\textbf{t}, \textbf{a}, \textbf{v}) \in \mathbb{R}^{3\times 3}$ is defined: 

\begin{equation}
    \textbf{G}(\textbf{t}, \textbf{a}, \textbf{v})=\textbf{A}^\top \textbf{A}= \begin{bmatrix}
 \left\langle \textbf{a},\textbf{a}\right\rangle& \left\langle \textbf{a},\textbf{v}\right\rangle & \left\langle \textbf{a},\textbf{t}\right\rangle \\
 \left\langle \textbf{v},\textbf{a}\right\rangle& \left\langle \textbf{v},\textbf{v}\right\rangle & \left\langle \textbf{v},\textbf{t}\right\rangle \\
 \left\langle \textbf{t},\textbf{a}\right\rangle& \left\langle \textbf{t},\textbf{v}\right\rangle & \left\langle \textbf{t},\textbf{t}\right\rangle
\end{bmatrix}.
\label{eq:gram}
\end{equation}

\noindent Notably, it has been shown that the determinant of the Gram matrix $\mathbf{G}$, also called the Gramian, is the square of the volume of the $3$-dimensional parallelotope formed by the vectors \cite{Gantmacher1959matrix}:

\begin{equation}
    \text{Vol}(\textbf{t}, \textbf{a}, \textbf{v})= \sqrt{\det   \mathbf{G}(\textbf{t}, \textbf{a}, \textbf{v})}.
    \label{eq:volume}
\end{equation}

The GRAM contrastive losses exploit the volume computation with the Gram matrix into the InfoNCE loss to align the three modalities at once:

\begin{equation}
\label{eq:contrastiveloss}
    \mathcal{L}_{AV2T}=-\frac{1}{B}\sum_{i=1}^{B}\log\frac{\exp(-\text{Vol}(\textbf{t}_i,\textbf{a}_{i},\textbf{v}_{i})/\tau)}{\sum_{j=1}^{K}\exp(-\text{Vol}(\textbf{t}_j,\textbf{a}_{i},\textbf{v}_{i})/\tau)},
\end{equation}
\begin{equation}
\label{eq:contrastiveloss2}
    \mathcal{L}_{T2AV}=-\frac{1}{B}\sum_{i=1}^{B}\log\frac{\exp(-\text{Vol}(\textbf{t}_i,\textbf{a}_{i},\textbf{v}_{i})/\tau)}{\sum_{j=1}^{K}\exp(-\text{Vol}(\textbf{t}_i,\textbf{a}_{j},\textbf{v}_{j})/\tau)},
\end{equation}

\noindent whereby $\tau$ is the temperature parameter and $B$ the batch size.

According to the GRAM loss function in \eqref{eq:contrastiveloss}, the GRAM model consists of three encoders to encode the different modalities into the latent space. The video modality is encoded with EVAClip-ViT-G \cite{Sun2023EVACLIPIT}, the text one with BERT-B \cite{Devlin2019BERTPO}, while the audio with BEATs \cite{beats2023}.

\subsection{Audio Synthesis Model}

Our audio synthesis model leverages Stable Audio Open \cite{Evans2024StableAO}, a state-of-the-art latent diffusion model (LDM) for generating high-quality, stereo audio at 44.1 kHz. 
While Stable Audio excels at generating semantically rich audio from text prompts, it lacks explicit mechanisms for temporal and multimodal conditioning, making it unsuitable for video-to-audio (V2A) tasks. 
To address this limitation, we introduce novel conditioning strategies leveraging the Gramian Representation Alignment Measure (GRAM) to guide the synthesis process semantically. 

The temporal alignment is provided using directly the envelope extracted with librosa library\footnote{\url{https://librosa.org/doc/main/generated/librosa.feature.rms.html}} from the ground truth audio, such as the main scope of this work is focusing on the semantic alignment and not introducing novel methods for temporal synchrony. 

\subsubsection{Semantic Control} our novel approach lies in the use of GRAM-aligned embeddings as conditioning inputs for the audio synthesis model. Unlike previous methods that rely on separately trained encoders (e.g., CLAP or CAVP) with unaligned latent spaces, our approach integrates GRAM-trained encoders to produce a unified latent representation for video, text, and audio modalities. This alignment ensures consistent and semantically meaningful interactions across modalities, enabling precise control over the audio generation process.
Specifically, we condition the audio synthesis model on a set of multimodal embeddings $\mathbf{F} = {\mathbf{f}_1, \mathbf{f}_2, \mathbf{f}_3}$, where each $\mathbf{f}_i$ represents a semantic embedding derived from GRAM encoders trained jointly across the three modalities. These embeddings are integrated into the diffusion process through cross-attention mechanisms, as originally proposed for global conditioning in Stable Audio \cite{Evans2024FastTL}.
During inference, we can use all the modality togheter, as done during training, or we can use them separately.

\subsubsection{Temporal Control} the temporal alignment is provided by an envelope extracted directly from the ground truth audio. 
The $i$-th sample of the temporal sequence representing the envelope is then calculated on a window of the audio signal $\mathbf{y}$ as follows:
\begin{equation}
    \mathbf{r}_i = \mathbf{RMS}_i(\mathbf{y)} = \sqrt{\frac{1}{W} \sum_{t=ih}^{ih+W}\mathbf{y}^{2}(t)},
\label{eqn:RMS}
\end{equation}
where $W$ is the window size and $h$ is the hop size. In our experiments we set $W=512$ and $h=128$.
The envelope serves as a coarse temporal guide, providing information about the timing and intensity of audio events. To encode this temporal control, we utilize the pre-trained VAE from Stable Audio, which downsamples the input stereo audio by a factor of 1024, mapping it into a compact latent space. The latent representation of the envelope, $\mathbf{r_c}$, is processed through a ControlNet-inspired architecture \cite{chen2024pixartalpha}, allowing fine-grained temporal adjustments during audio generation.

\subsubsection{Diffusion Process} our audio model is based on Stable Audio and follows the standard latent denoising diffusion formulation. Given a noisy latent representation $\mathbf{z} = \mathcal{E}(\mathbf{y})$ at time step $t$, the model learns to estimate the noise $\epsilon_{\theta}(\mathbf{z}_t, t, \mathbf{F}, \mathbf{r_c})$ conditioned on semantic embeddings $\mathbf{F}$ and the temporal control signal $\mathbf{r_c}$. 
In the forward process, Gaussian noise is slowly added to the original data distribution with a fixed schedule $\alpha_1, \ldots, \alpha_T$, where $T$ is the total timesteps, and $\bar{\alpha}_t = \prod_{i=1}^{t} \alpha_i$:

\begin{equation}
q(\mathbf{z}_t | \mathbf{z}_{t-1}) = \mathcal{N}(\mathbf{z}_t; \sqrt{\alpha_t} \mathbf{z}_{t-1}, (1 - \alpha_t) \mathbf{I})
\end{equation}
\begin{equation}
q(\mathbf{z}_t | \mathbf{z}_0) = \mathcal{N}(\mathbf{z}_t; \sqrt{\bar{\alpha}_t} \mathbf{z}_0, (1 - \bar{\alpha}_t) \mathbf{I}).
\end{equation}

The training objective is the the same L2 loss on which Stable Audio
models are trained\cite{Evans2024LongformMG}. 
%minimizes the weighted sum of:
%\begin{itemize}
%\item Reconstruction Loss: a multiresolution STFT loss computed on mid-side (M/S) and left-right (L/R) stereo representations to ensure perceptual fidelity.

%\item Adversarial Loss: feature matching loss leveraging convolutional discriminators to enhance audio realism.

%\item KL Divergence: regularizes the latent space to match the prior distribution, weighted by $10^{-4}$.
%\end{itemize}
After training, LDMs generate latents by sampling through the reverse process with $\mathbf{z}_T \sim \mathcal{N}(0, \mathbf{I})$ formulated as:

\begin{equation}
p_\theta(\mathbf{z}_{t-1} | \mathbf{z}_t) = \mathcal{N}(\mathbf{z}_{t-1}; \mu_\theta(\mathbf{z}_t, t, \mathbf{F}, \mathbf{r_c}), \sigma_t^2 \mathbf{I})
\end{equation}

\begin{equation}
\mu_\theta(\mathbf{z}_t, t, \mathbf{F}, \mathbf{r_c}) = \frac{1}{\sqrt{\alpha_t}} \left( \mathbf{z}_t - \frac{1 - \alpha_t}{\sqrt{1 - \bar{\alpha}_t}} \epsilon_\theta(\mathbf{z}_t, t,  \mathbf{F}, \mathbf{r_c}) \right)
\end{equation}

\begin{equation}
\sigma_t^2 = \frac{1 - \bar{\alpha}_{t-1}}{1 - \bar{\alpha}_t} (1 - \alpha_t).
\end{equation}

Finally, the desired output  $\hat{\mathbf{y}}$ is obtained by decoding the generated latent $\mathbf{z}_0$ with a decoder $\mathcal{D}$.

We freeze the pre-trained weights of the diffusion model and only train the ControlNet layers, which process the RMS envelope, and the linear projections that align GRAM embeddings to the conditioning dimensions of Stable Audio. By jointly leveraging GRAM-aligned embeddings for semantic control and the ControlNet mechanism for temporal alignment, our model ensures that the generated audio aligns both semantically and temporally with the input video. A block diagram of the proposed architecture is shown in Fig~\ref{fig:architecture}.

The ControlNet model is trained with the v-prediction MSE loss $\mathcal{L} = \mathbb{E}[v_\theta(\mathbf{z_t}, t, \mathbf{r_c}) - v]$, where $v = \sqrt{\bar{\alpha_t}}\epsilon - \sqrt{1-\bar{\alpha_t}} x_0$.

\section{Experiments}
\label{sec:exp}
\subsection{Dataset}

We work with the \textit{Greatest Hits} dataset \cite{Owens2015VisuallyIS}, a well-known and widespread benchmark for video-to-audio generation tasks. The dataset contains videos of people using a drumstick to strike or rub different surfaces and objects. The choice of a drumstick as main object in motion allows the scene’s action to remain clearly visible with minimal occlusion of the frame. Each video captures the sound of these interactions with a shotgun microphone attached to the camera, and the audio is later processed to remove noise. Metadata provided for each video is used to create textual prompts following the structure proposed in Fol·AI \cite{gramaccioni2024folai}: “A person \textit{\{action\}} \textit{\{frequency\}} on \textit{\{material\}} with a wooden stick.” The placeholders \textit{\{action\}} (e.g., “hit” or “scratch”), \textit{\{frequency\}} (e.g., “multiple times” or “once”), and \textit{\{material\}} are populated based on the metadata details.
The carefully curated samples of this dataset are crucial for V2A model training, as real-world video datasets often lack both the audiovisual alignment and the quality required to make models understand how to produce audio that is semantically and temporally consistent with the input video. This dataset contains 977 video recordings captured in diverse settings, both indoors and outdoors. Indoor videos showcase materials like metal, plastic, and cloth, while outdoor recordings feature dynamic materials such as water, leaves, and grass. We extract 10-second-long chunks from each sample to train and test our model. On average, each video includes 48 distinct actions, split between striking and rubbing, ensuring that each extracted chunk has sufficient activity.  
For our experiments, we split the dataset into 732 videos for training, 49 for validation, and 196 for testing.

\subsection{Evaluation Metrics}

For an objective evaluation of our model, we utilize the most commonly adopted metrics to assess semantic quality in V2A tasks:
\begin{itemize}
\item Fréchet Audio Distance (FAD):
FAD \cite{KilgourZRS19} is a metric designed to assess the quality and realism of generated audio by comparing it to reference audio. It evaluates the similarity between the statistical distributions of embeddings extracted from real and generated waveforms. The choice of the audio encoder for extracting these embeddings plays a crucial role, as different encoders emphasize various audio features, affecting how well the metric aligns with human perception \cite{Tailleur2024CorrelationOF}. To account for this, we calculate FAD using two distinct audio encoders: Microsoft CLAP (FAD-C) \cite{Elizalde2023CLAPLA}, and Laion-CLAP (FAD-LC) \cite{laionclap2023}. The FAD scores are computed using the \textit{fadtk}\footnote{\url{https://github.com/DCASE2024-Task7-Sound-Scene-Synthesis/fadtk}} library.

\item CLAP-score:
The CLAP-score evaluates the overall quality of the generated waveforms, also used in \cite{Evans2024StableAO}. It calculates the cosine similarity between embeddings of ground truth and generated audio, which are obtained using the CLAP model \cite{laionclap2023}. Given that the majority of baseline models employs CLAP as the primary audio representation, this metric serves as a key indicator of how effectively the conditioning features contribute to generating the final output for a fair comparison.

\item Fréchet Audio-Visual Distance (FAVD):
FAVD \cite{10.1007/978-3-031-72986-7_17} is increasingly recognized for evaluating video-to-audio (V2A) models. It measures the alignment, both temporal and semantic, between the audio and video modalities. This metric calculates the Fréchet Distance between video embeddings and audio embeddings. For our evaluation, we use I3D \cite{Carreira2017QuoVA} as the video encoder and VGGish \cite{Hershey2016CNNAF} as the audio encoder, extracting embeddings from both ground truth videos and the generated audio to determine their alignment.

%\item GRAM: The Gramian Reperesenation Alignment Measure (GRAM) has been proven to be a good proxy for representation alignment in the latent space \cite{cicchetti2024GRAM}. In particular, the GRAM provides unique insights on multiple modalities such as video, text, and audio, alignment in a joint fashion, without requiring pairwise comparisons. In our experiments, GRAM measures the latent alignment of multiple sources (video and text) and of the generated audio, providing information on how close the generated audio is to the original video. GRAM computes the volume of the parallelotope defined by the $n$ embeddings in input. Smaller volume means representation are highly aligned, while larger volume stands for misaligned representations. 
\end{itemize}
\subsection{Training and Inference Details} 

For training FoleyGRAM, we initialize the model weights using the Stable Audio Open repository and its associated checkpoint. The ground truth audio used in our experiments is 44.1 kHz stereo recordings from the Greatest Hits dataset. The model is trained on a single Nvidia RTX A6000 GPU (48 GB) with a batch size of 12 for 20,000 steps. The training process employs the AdamW optimizer, with parameters configured as those in Stable Audio Open, and uses a fixed learning rate of \( 1 \times 10^{-4} \).  

To initialize GRAM encoders, we use the official associated repository and its relative checkpoints. Three GRAM encoders, which are EVAClip-ViT-G for video, BEATs for audio, and BERT-B for text with a total number of parameters equal to 1B, have been previously pretrained on the VAST27M dataset \cite{Chen2023VASTAV} with conventional contrastive loss functions. Later, the learned latent space is rearranged and pretrained on a subset of such dataset comprising 150k samples with the GRAM losses in \eqref{eq:contrastiveloss} and \eqref{eq:contrastiveloss2}, and finally fine-tuned on the Greatest Hits dataset to make the encoders aware of the particular cases of this dataset. The pertaining on the subset of VAST27M dataset has been carried on for one epoch with learning rate $ 1 \times 10^{-4}$ with a batch size of 256 on 4 NVIDIA A100 cards, and the same configuration holds for the fine-tuning on Greatest Hits.

During inference, envelopes extracted directly from ground truth audio are interpolated to match the target sample rate, and fed into the ControlNet from the audio synthesis model as inputs. The model then generates the final output in 150 sampling steps, applying classifier-free guidance with a guidance scale set to 2.

\begin{table*}[!t]
\centering
\caption{Results for FoleyGRAM and comparison with other SOTA models on \textit{Greatest Hits}. HRC stands for Human Readable Control and refers to the use of time conditionings signals that sound designers can use to control the generation process (i.e., envelope or onsets). Our model provides the best results on all objective metrics compared to the baselines.}
\label{tab:overall-results}
\begin{tabular}{l|c|ccccc}
\toprule
 \textbf{Model}  & \textbf{HRC}  & \textbf{FAD-C ↓} & \textbf{FAD-LC ↓}  & \textbf{CLAP ↑} & \textbf{FAVD ↓} \\
\midrule
    SpecVQGAN \cite{SpecVQGAN_Iashin_2021}  & \ding{55}   & 1001 & 0.7102 & 0.1418 & 0.1418 \\
    Diff-Foley \cite{NEURIPS2023_98c50f47}  & \ding{55} & 654 & 0.4690 & 0.3733 & 0.3733 & \\
    CondFoleyGen \cite{du2023conditional}  & \ding{55}  & 650 & 0.4883 & 0.4879 & 0.4879 \\
    
\midrule
    SyncFusion \cite{Comunit2023SyncfusionMO} (Audio)
     & \ding{51}  & 591 & 0.4365 & 0.5154 & 0.5154 \\
    SyncFusion (Text) & \ding{51} & 542 & 0.2793 & 0.6621 & 0.6621 \\
    
\midrule
    Video-Foley \cite{Lee2024VideoFoleyTV} (Audio)  & \ding{51}  & 644 & 0.4997 & 0.3680 & 0.3680 \\
    Video-Foley (Text)  & \ding{51}  & 435 & 0.1671 & 0.6779 & 0.6779 \\

\midrule
    FoleyGRAM (Ours)
     & \ding{51}  & \textbf{235} & \textbf{0.0720} & \textbf{0.7083} & \textbf{0.8912} \\
\bottomrule
\end{tabular}
\end{table*}

\begin{table*}[!t]
\centering
    \caption{Ablation studies: conditioning FoleyGRAM with all modalities (AVT), audio and video (AV), audio and text (AT), video and text (VT), audio (A), video (V) and text (T) modalities. For all the experiments, the conditioning modalities are the ground truth samples, even though at inference time any kind of sample can be used to condition the semantics of the waveforms.}
    \centering
    \begin{tabular}{ccccc}
        \toprule
        \textbf{Conditionings} & \textbf{FAD-C $\downarrow$} & \textbf{FAD-LC $\downarrow$} & \textbf{FAVD $\downarrow$} & \textbf{CLAP $\uparrow$} \\
        \midrule
        AVT  & \textbf{235}  & \textbf{0.072}  & \textbf{0.8912}  & \textbf{0.7083} \\
        AV   & 238  & 0.074  & 0.9309  & 0.7007 \\
        AT   & 287  & 0.093  & 0.9978  & 0.6814 \\
        VT   & 269  & 0.119  & 1.1739  & 0.6623 \\
        A    & 325  & 0.135  & 1.6513  & 0.6155 \\
        V    & 271  & 0.122  & 1.2003  & 0.6543 \\
        T    & 1069 & 0.797  & 6.1288       & 0.1962 \\
        \bottomrule
    \end{tabular}

    \label{tab:ablation_results}
\end{table*}

\section{Results}
\label{sec:res}

\subsection{Baselines}
We evaluate our model against the main publicly available V2A models at the time of this study.
\subsubsection{SpecVQGAN} it extracts RGB and optical flow features of a video and leverages a Transformer-based autoregressive architecture to generate temporally and semantically aligned audio to the reference video.
\subsubsection{CondFoleyGen} this model uses a similar architecture respect to SpecVQGAN, adding additional controls on the final output conditioning with audio and video features from the semantic target. The model is trained directly using Greatest Hits, succeeding in achieving an efficient alignment in both content and
timing with the reference video.
\subsubsection{Diff-Foley} leverages Contrastive Audio-Visual Pretraining (CAVP) to achieve temporal and semantic alignment between audio and video modalities, enabling the generation of video embeddings with features pertinent to the associated audio. These embeddings are then employed as direct conditioning inputs for Stable-Diffusion.
\subsubsection{SyncFusion} this model is the first to introduce a human-readable control mechanism for the V2A task. It utilizes a ResNet(2+1)D-18 based video encoder, which processes video frames to generate an onset track. This onset track is subsequently fed into a time-domain diffusion model to produce the final audio output. 
%The onset track serves as a visual representation of the annotations sound designers use to determine the timing of sound sources in a video, offering users a clear and editable control over the audio generation process. Represented as a binary mask, the onset track indicates whether the action of interest occurs in each video frame, providing temporal placement information for sounds but lacking details about intensity or duration.
\subsubsection{Video-Foley} this model uses a video encoder through which the RMS of the audio signal associated with the input video can be mapped, which is then used as the control signal for the temporal alignment of the model. In contrast, the semantics of the final output is controlled by embeddings produced by the CLAP audio/text encoder. These control signals are used to generate 16kHz mono audio through the use of AudioLDM.

For all of the above models, we use the official released codes provided on GitHub and relative checkpoints. 

\subsection{Discussion}
As shown in Table \ref{tab:overall-results}, FoleyGRAM demonstrates substantial improvements in semantic quality of the generated audio compared to all baseline models. This enhancement is primarily attributed to the integration of the multimodal-aligned encoder GRAM for conditioning the state-of-the-art audio generation model, Stable Audio. Unlike Video-Foley and SyncFusion, which rely on CLAP as the audio encoder for semantic conditioning, our approach leverages GRAM to ensure alignment across audio, video, and text modalities. This alignment enables FoleyGRAM to better capture the semantic features required for precise audio generation.
Our model is also able to provide strong results on CLAP-based metrics, surpassing even Video-Foley and SyncFusion, despite the latter directly rely on CLAP for their semantic encoders.
The improved evaluation metrics scores of FoleyGRAM confirm the advantages of employing a unified multimodal encoder like GRAM for conditioning, particularly in scenarios where cross-modal consistency is essential. 
Additionally, the use of Stable Audio as the backbone for audio generation ensures high-definition, stereo audio at 44.1 kHz, aligning with professional audio standards. The integration of ControlNet within our architecture further enhances the ability of the model to incorporate temporal conditioning through envelopes, ensuring precise timing and dynamic for the generated waveforms. Notably, FoleyGRAM achieves these results while being lightweight and efficient, requiring only approximately six hours of data of which the Greatest Hits dataset is composed and a limited number of training steps. This efficiency underscores the robustness and practicality of our approach for real-world sound design applications.

\subsection{Ablation studies} 
GRAM allows the alignment of three modalities, audio video and text, which can be used together to provide meaningful semantic information to the synthesis model. Conditioning with multiple modalities allows for better control over the semantics of the generated waveform. To demonstrate this assertion also in our generation task, at inference time we conditioned our model in seven different ways: first using all three modalities simultaneously (AVT), then audio and video (AV), audio and text (AT) and video and text (VT), and finally the single modalities audio (A), Video (V) and text (T). The results shown in Table \ref{tab:ablation_results} demonstrate that using multiple modalities simultaneously succeeds in providing the model with more semantic information, achieving the best results in the case of AVT conditioning. 

% Resukts are provided in Table \ref{tab:ablation_results}
% \label{sec-F:exp}

\label{sec-F:exp}

\section{Conclusion}
\label{sec:con}
In this paper, we introduced FoleyGRAM, a novel V2A synthesis model that combines a state-of-the-art audio generation framework, Stable Audio, with GRAM, a unified multimodal encoder designed for cross-modal alignment. Our results demonstrate significant advancements in the semantic accuracy, achieving strong performances across key semantic metrics. By leveraging GRAM as the primary encoder for semantic conditioning, FoleyGRAM can use multiple aligned modalities simultaneously in order to leverage as much information as possible to generate waveforms with rich semantic information. Additionally, the integration of ControlNet allows for precise temporal control through envelopes, enabling the generation of high-quality 44.1 kHz stereo audio. 
FoleyGRAM achieves these results with a lightweight architecture and efficient training, requiring only a small dataset and limited computational resources. This makes our model a powerful tool for sound designers and also a practical solution for real-world applications where resource constraints are a factor. 
The proposed model wants to encourage further exploration of multimodal deep learning in V2A tasks, highlighting the potential of unified embeddings and advanced generative models to bridge the gap between visual and audio modalities.

\section{Acknowledgements}
This work was supported by the European Union under the Italian National Recovery and Resilience Plan (NRRP) of NextGenerationEU, partnership on “National Centre for HPC, Big Data and Quantum Computing” (CN00000013 - Spoke 6: Multiscale Modelling \& Engineering Applications). 

This work was supported by “Progetti di Ricerca Medi” of Sapienza University of Rome for the project “SAID: Solving Audio Inverse problems with Diffusion models”, under grant number RM123188F75F8072.

\bibliography{ref}
\bibliographystyle{ieeetr}
\end{document}